\documentclass[A4,11pt]{article}
\usepackage{amsfonts}
\usepackage{amsmath,amsthm}
\usepackage{amssymb,amscd}
\usepackage{blkarray}
\usepackage{enumerate}
\usepackage{kbordermatrix}
\usepackage{color}
\usepackage{cite}
\usepackage{esint}
\usepackage{tikz}
\usepackage{tikzscale}
\usepackage[absolute,overlay]{textpos}
\usepackage{multicol}
\usepackage{framed}
\usepackage{lscape}
\usepackage{caption}
\usepackage{subcaption}
\usepackage[inline]{enumitem}
\usepackage{float}
\usepackage{mathtools}
\usepackage{lipsum}
\renewcommand{\kbldelim}{(}
\renewcommand{\kbrdelim}{)}
\makeatletter
\newcommand{\rmnum}[1]{\romannumeral #1}
\newcommand{\Rmnum}[1]{\expandafter\@slowromancap\romannumeral #1@}
\makeatother
\def\Xint#1{\mathchoice
{\XXint\displaystyle\textstyle{#1}}%
{\XXint\textstyle\scriptstyle{#1}}%
{\XXint\scriptstyle\scriptscriptstyle{#1}}%
{\XXint\scriptscriptstyle\scriptscriptstyle{#1}}%
\!\int}
\def\XXint#1#2#3{{\setbox0=\hbox{$#1{#2#3}{\int}$}
\vcenter{\hbox{$#2#3$}}\kern-.5\wd0}}

\def\dashint{\Xint-}

\def\Xint#1{\mathchoice
{\XXint\displaystyle\textstyle{#1}}%
{\XXint\textstyle\scriptstyle{#1}}%
{\XXint\scriptstyle\scriptscriptstyle{#1}}%
{\XXint\scriptscriptstyle\scriptscriptstyle{#1}}%
\!\int}
\def\XXint#1#2#3{{\setbox0=\hbox{$#1{#2#3}{\int}$}
\vcenter{\hbox{$#2#3$}}\kern-.5\wd0}}

\def\dashint{\Xint-}

\theoremstyle{definition}
\newtheorem{theorem}{Theorem}[section]

\newtheorem{definition}{Definition}[section]

\newtheorem{proposition}[theorem]{Proposition}

\theoremstyle{remark}
\newtheorem{remark}{Remark}[section]

\begin{document}
\title{Integrable multispecies totally asymmetric stochastic interacting particle systems with homogeneous rates}
\author{\textbf{Eunghyun Lee\footnote{eunghyun.lee@nu.edu.kz}~ and Temirlan Raimbekov \footnote{temirlan.raimbekov@alumni.nu.edu.kz}}\\ {\text{Department of Mathematics,}}
                                         \date{}   \\ {\text{School of Sciences and Humanities,}} \\ {\text{Nazarbayev University, }}\\ {\text{Kazakhstan }}   }

\date{}
\maketitle
\begin{abstract}
\noindent We study one-dimensional stochastic particle systems with exclusion interaction—each site can be occupied by at most one particle—and homogeneous jumping rates. Earlier work of Alimohammadi and Ahmadi classified 28 Yang–Baxter integrable two-particle interaction rules for two-species models with homogeneous rates (see Section \ref{Introduction} for precise references). In this work, we show that 7 of these 28 cases can be naturally extended to integrable models with an arbitrary number of species $N \geq 2$. A key novelty of our approach is the discovery of new integrable families with one or two continuous parameters that generalize these 7 cases, significantly broadening the known class of multi-species integrable exclusion processes. Furthermore, for 8 of the remaining 21 cases, we propose an alternative extension scheme that also yields integrable $N$-species models, thereby opening new directions for constructing and classifying integrable particle systems.
\end{abstract}
\textbf{Key words}:
Exactly solvable models, Bethe ansatz, Yang-Baxter equation, Markov chain

\section{Introduction}\label{Introduction}
Over the past two decades, the study of nonequilibrium statistical mechanics has seen significant progress through the rise of integrable probability, a field that merges exactly solvable models with modern probabilistic and combinatorial methods. These developments have deepened our understanding of complex systems far from equilibrium, especially in one dimension, where fluctuations and correlations are markedly different from their equilibrium counterparts. Central to this progress have been models such as the asymmetric simple exclusion process (ASEP), its various generalizations, and the six-vertex model, which were systematically developed in foundational works on interacting particle systems and stochastic growth \cite{Liggett1,Liggett2,Spohn,Schutz-2001}, and further advanced in probabilistic and integrable frameworks \cite{Borodin-Corwin,Borodin-Corwin-Petrov-Sasamoto,Corwin}.

A defining feature of integrable models is the presence of exact solvability through algebraic structures such as the Yang–Baxter equation, the Bethe ansatz, and quantum group symmetries, which allow one to compute observables exactly despite the stochastic and interacting nature of the systems.Seminal contributions include exact transition probability formulas for ASEP \cite{Schutz-1997,Tracy-Widom-2008}, which later provided a key foundation for rigorous results in the Kardar–Parisi–Zhang (KPZ) universality class \cite{Amir,KPZ,Nagao,TW3}. The Bethe ansatz, in particular, has proven to be a powerful method not only for exclusion processes but also for systems with unbounded site occupancy, leading to a variety of exactly solvable zero-range and misanthrope-type processes \cite{Korhonen-Lee-2014,Lee-2012,Lee-Wang-2017,Povol,Povol2,Povolotsky-2,Wang-Waugh}.

In recent years, multi-species generalizations of these models have attracted considerable attention. These systems, which incorporate interactions between distinguishable particle types, exhibit rich algebraic structures and complex dynamical behavior. They have been studied from perspectives ranging from integrable probability \cite{Borodin-Bufetov,Aggarwal-Nicoletti-Petrov,Kuan-2020,Gier-Mead-Wheeler-2021} to representation-theoretic and combinatorial approaches \cite{Ayyer-Kuniba,Ayyer-Martin,Kuniba-Maruyama-Okado,Kuniba-Okado-Watanabe}, as well as through analytic treatments of multi-species ASEP and related exclusion processes \cite{Tracy-Widom-2009,Tracy-Widom-2013}. Notably, Yang–Baxter integrability has played a central role in analyzing such systems. In addition to these more recent developments, several earlier contributions also deserve attention, in particular works on two-species exclusion and their integrability properties \cite{Aghamohammadi-Fatollahi-Khorrami-Shariati,Ali,Ali2,Ali-Ahmadi,Roshani-Khorrami2,Roshani-Khorrami}.

In this paper, we further investigate and extend the work of Alimohammadi and Ahmadi \cite{Ali-Ahmadi}. Consider the one-dimensional integer lattice $\mathbb{Z}$, where $n$ particles move to the right according to specified rules. Each particle at $x$ is labeled by a positive integer in $\{1,\dots, N\}$, referred to as its species, and independently attempts to jump to the neighboring site on the right, that is, $x+1$, after an exponential waiting time with rate $1$. The system follows the exclusion rule: each site can be occupied by at most one particle. The precise dynamics of the model are determined by the interaction rule that determines what happens when a particle attempts to jump to an already occupied site.

A state of the system is represented by a pair $(X,\pi)$ or equivalently $(x_1,\dots,x_n, \pi_1\cdots\pi_n)$, where $X=(x_1,\dots, x_n)\in \mathbb{Z}^n$  denotes the positions of the particles with $x_1< \cdots< x_n$, so that $x_i$ is the position of the $i$th leftmost particle, and  $\pi = \pi_1\pi_2\cdots \pi_n$ is a word of length $n$ with entries in $\{1,\dots, N\}$, indicating that $\pi_i$ is the species of the $i$th leftmost particle.

One of the simplest models is the totally asymmetric simple exclusion process (TASEP), where there is only one species, jumps occur only to the right, and any attempted jump to an occupied site is simply blocked. In the multispecies version of the TASEP, which is called the multispecies TASEP (mTASEP), the interaction rule is as follows.
\begin{quote}
If a particle of species $i$ with $i>j$ attempts to jump to a site occupied by a particle of species $j$, the two particles exchange positions. If $i \leq j$, the jump is blocked.
\end{quote}
 This  interaction rule  can be equivalently interpreted by introducing \textit{intermediate configurations}  in which a site temporarily accommodates two particles. These configurations arise instantaneously during interactions and vanish immediately afterward. Since they do not belong to the actual state space, the exclusion rule remains intact.
We will refer to such a temporary configuration as \textit{a hidden state}. If site $x$ temporarily holds two particles of species $i$ and $j$, with $i$ occupying  the left position and  $j$ the right position at $x$, we denote this hidden state  by $(x,x,ij)$.

Then, the dynamics of the mTASEP can be  described as follows.
\begin{quote}
Each particle of species $i$ at site $x$ independently jumps to site $x+1$ after an exponential waiting time with rate $1$.  If $x+1$ is occupied by a particle of species  $j$, then the site $x+1$ temporarily accommodates both particles,  with the particle of species $\min(i,j)$ occupying the left position and  the particle of species $\max(i,j)$ the right position at $x+1$. Immediately thereafter, the particle of species $\min(i,j)$ moves back to site $x$, resulting in an exchange of positions if $i>j$, or no net movement if $i\leq j$.
\end{quote}
\begin{figure}[H]
  \centering
  \begin{tikzpicture}

    \draw[draw=black] (2.5,0.5) rectangle (8.5,-0.5);

    \node[circle, draw, minimum size=0.4cm, inner sep = 0] (left1) at (0,0) {$1$};
    \node[circle, draw, minimum size=0.4cm, inner sep = 0] (left2) at (1,0) {$1$};

    \node[circle, draw, dashed, thick, minimum size=0.4cm, inner sep = 0] (mid1) at (3.1,0) {};
    \node[circle, draw, minimum size=0.4cm, inner sep = 0] (mid2) at (4.1,0) {$1$};
    \node[circle, draw, minimum size=0.4cm, inner sep = 0] at (4.5,0) {$1$};

    \node[circle, draw, dashed, thick, minimum size=0.4cm, inner sep = 0] at (6.5,0) {};
    \node[circle, draw, minimum size=0.4cm, inner sep = 0] at (7.5,0) {$1$};
    \node[circle, draw, minimum size=0.4cm, inner sep = 0] at (7.9,0) {$1$};

    \node[circle, draw, minimum size=0.4cm, inner sep = 0] at (10.0,0) {$1$};
    \node[circle, draw, minimum size=0.4cm, inner sep = 0] at (11.0,0) {$1$};

    \draw[->] (1.8,0) -- (2.2,0);
    \draw[->, dash pattern=on 2pt off 1.5pt] (5.3,0) -- (5.7,0);
    \draw[->, dash pattern=on 2pt off 1.5pt] (8.8,0) -- (9.2,0);
  \end{tikzpicture}

  \bigskip

  \begin{tikzpicture}

    \draw[draw=black] (2.5,0.5) rectangle (8.5,-0.5);

    \node[circle, draw, minimum size=0.4cm, inner sep = 0] (left1) at (0,0) {$2$};
    \node[circle, draw, minimum size=0.4cm, inner sep = 0] (left2) at (1,0) {$1$};

    \node[circle, draw, dashed, thick, minimum size=0.4cm, inner sep = 0] (mid1) at (3.1,0) {};
    \node[circle, draw, minimum size=0.4cm, inner sep = 0] (mid2) at (4.1,0) {$2$};
    \node[circle, draw, minimum size=0.4cm, inner sep = 0] at (4.5,0) {$1$};

    \node[circle, draw, dashed, thick, minimum size=0.4cm, inner sep = 0] at (6.5,0) {};
    \node[circle, draw, minimum size=0.4cm, inner sep = 0] at (7.5,0) {$1$};
    \node[circle, draw, minimum size=0.4cm, inner sep = 0] at (7.9,0) {$2$};

    \node[circle, draw, minimum size=0.4cm, inner sep = 0] at (10.0,0) {$1$};
    \node[circle, draw, minimum size=0.4cm, inner sep = 0] at (11.0,0) {$2$};

    \draw[->] (1.8,0) -- (2.2,0);
    \draw[->, dash pattern=on 2pt off 1.5pt] (5.3,0) -- (5.7,0);
    \draw[->, dash pattern=on 2pt off 1.5pt] (8.8,0) -- (9.2,0);
  \end{tikzpicture}

  \bigskip

  \begin{tikzpicture}

    \draw[draw=black] (2.5,0.5) rectangle (8.5,-0.5);

    \node[circle, draw, minimum size=0.4cm, inner sep = 0] (left1) at (0,0) {$1$};
    \node[circle, draw, minimum size=0.4cm, inner sep = 0] (left2) at (1,0) {$2$};

    \node[circle, draw, dashed, thick, minimum size=0.4cm, inner sep = 0] (mid1) at (3.1,0) {};
    \node[circle, draw, minimum size=0.4cm, inner sep = 0] (mid2) at (4.1,0) {$1$};
    \node[circle, draw, minimum size=0.4cm, inner sep = 0] at (4.5,0) {$2$};

    \node[circle, draw, dashed, thick, minimum size=0.4cm, inner sep = 0] at (6.5,0) {};
    \node[circle, draw, minimum size=0.4cm, inner sep = 0] at (7.5,0) {$1$};
    \node[circle, draw, minimum size=0.4cm, inner sep = 0] at (7.9,0) {$2$};

    \node[circle, draw, minimum size=0.4cm, inner sep = 0] at (10.0,0) {$1$};
    \node[circle, draw, minimum size=0.4cm, inner sep = 0] at (11.0,0) {$2$};

    \draw[->] (1.8,0) -- (2.2,0);
    \draw[->, dash pattern=on 2pt off 1.5pt] (5.3,0) -- (5.7,0);
    \draw[->, dash pattern=on 2pt off 1.5pt] (8.8,0) -- (9.2,0);
  \end{tikzpicture}

  \caption{The transitions $\dashrightarrow$ occur immediately.}
  \label{fig:fig1}
\end{figure}
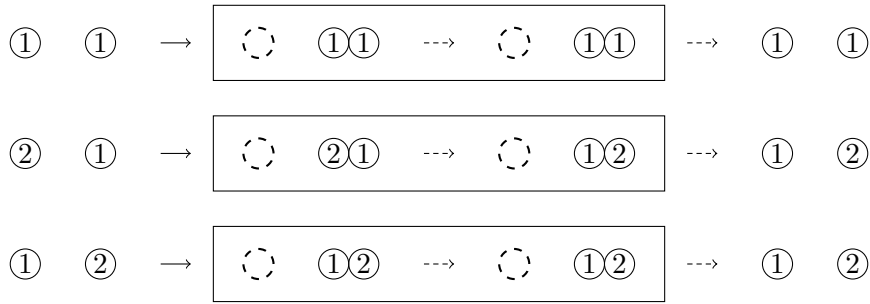
Figure \ref{fig:fig1} illustrates the two-particle interaction rule in the mTASEP that  consists of two steps. First, temporary placement of two particles at the same site in a prescribed order (the transitions in boxes in Figure \ref{fig:fig1}). Second, moving the particle of $\min(i,j)$ backward - against the direction of movement of the system - to complete the interaction.
The transitions in boxes in Figure \ref{fig:fig1} can be encoded by  a $4\times 4$ matrix whose rows and columns are labelled by the ordered pairs $11,12,21,22$ corresponding to the species of two particle,
 \begin{equation}\label{623pm720}
\kbordermatrix{
    & 11 & 12 & 21 & 22  \\
    11 & 1 & 0 & 0 & 0  \\
    12 & 0 & 1 & 1 & 0  \\
    21 & 0 & 0 & 0 & 0  \\
    22 & 0 & 0 & 0 & 1
  },
\end{equation}
where $(ij,kl)$-entry represents the probability of the transition $(x+1,x+1,kl) \dashrightarrow (x+1,x+1,ij)$ occurring. While the matrix (\ref{623pm720}) describes the interaction between particles of species $1$ and $2$, one can assume that the same matrix  governs interactions between particles of species $i$ and $j$ for any all $i<j$. Under this assumption, it is well known that  the corresponding scattering matrix (S-matrix)  satisfies the Yang-Baxter equation. This confirms the integrability of the multispecies TASEP. Moreover,  the explicit formula for its transition probabilities has been derived in \cite{Lee-2020}.

A natural question then arises: can a stochastic particle model defined by  a different two-particle interaction rule, that is, a different $4\times 4$ matrix than (\ref{623pm720}) be also integrable? This question was investigated by Alimohammadi and Ahmadi \cite{Ali-Ahmadi} for totally asymmetric two-species models with the following properties: (i) the particles jump to the right neighboring sites if these sites are not occupied (they did not consider a concept of hidden states), (ii) the interaction occurs only between nearest-neighbor particles, (iii) the particles can be annihilated or created, with the only restriction being that the total number of particles is conserved, and (iv) all the interactions occur with the same rate.  They identified all possible 28 physically distinct integrable totally asymmetric two-species models, including the simple diffusion process not distinguishing species.
\begin{remark}\label{926pm82}
Appendix \ref{627pm729} lists all 28 matrices obtained by Alimohammadi and Ahmadi in \cite{Ali-Ahmadi} up to the symmetry under interchange of species $1$ and $2$. The matrix (\ref{623pm720}) is physically equivalent to $\mathbf{b}^{(2)}$ in Appendix \ref{627pm729} by this symmetry. We will use the notation $\mathbf{\overline{b}}^{(l)}$ to denote the matrix obtained by interchanging the species labels $1\leftrightarrow 2$ in  the row and column indices of $\mathbf{b}^{(l)}$.
\end{remark}

Although the analytical tools we use are standard, the novelty of this paper lies in the new multi-species integrable models and extension schemes that we establish.

\subsection*{Summary of the main results}

This paper builds directly on the classification of Alimohammadi and Ahmadi \cite{Ali-Ahmadi}, who identified 28 Yang--Baxter integrable two-species interaction rules. A natural expectation might be that if a two-species model is integrable, then its $N$-species extension should also be integrable for all $N \geq 2$. Our findings show that this expectation is not always correct: among the 28 cases, only 7 admit natural extensions to integrable $N$-species models.

The originality of this work lies in clarifying these limitations and, at the same time, in expanding the class of known integrable multi-species systems. In particular, we establish the following main results.
\begin{itemize}
  \item We show that only 7 of the 28 two-species models classified in \cite{Ali-Ahmadi} extend naturally to integrable $N$-species systems.
  \item We construct new one- and two-parameter families of integrable models, which broaden and generalize these 7 cases.
  \item For the remaining 21 cases, we introduce an alternative extension scheme that yields 8 further integrable $N$-species models.
  \item We provide explicit formulas for transition probabilities and demonstrate that all of the above results remain valid when the dynamics incorporate the drop--push interaction rule.
\end{itemize}

Compared to the work of Alimohammadi and Ahmadi, which was limited to a two-species classification, our results both pinpoint the exact subset of models that extend naturally and develop new constructions that significantly enlarge the family of known integrable multi-species systems. These contributions underscore the novelty and significance of the present paper.

This paper is organized as follows. In Section \ref{prelim}, we introduce notations, define the natural extension, and review some background on tensor products. Section \ref{main} presents our main results, and Section \ref{discussion} discusses the integrability of the models with the drop-push interaction rule and the transition probability formulas. In Appendix \ref{627pm729}, we include the complete list of integrable $4\times 4$ matrices for the two-species case identified by Alimohammadi and Ahmadi.
\section{Preliminary}\label{prelim}
\subsection{Notations}
We introduce some notations that will be used throughout the paper, following the conventions adopted in the authors' previous works \cite{Lee-2020,Lee-2024}. Let $P_{(Y,\nu)}(X,\pi;t)$ be the transition probability from state $(Y,\nu)$ to state $(X,\pi)$ over a time interval of length $t$, where $X=(x_1,\dots, x_n)$ and $Y=(y_1,\dots, y_n)$ are ordered configurations with $x_1<\cdots <x_n$ and $y_1< \cdots <y_n$, and $\pi=\pi_1\cdots \pi_n$ and $\nu=\nu_1\cdots \nu_n$ are  words of length $n$ consisting of elements in $\{1,\dots, N\}$, that is, they are  permutations of certain multisets.

Let $\mathbf{P}(t)$ be the infinite matrix whose columns are indexed by  initial states $(Y,\nu)$ and rows by terminal states $(X,\pi)$  after time $t$, with all admissible $X$ and $Y$ and words $\pi, \nu$ ranging from $1\cdots 1$ to $N\cdots N$.  The $((X,\pi),(Y,\nu))$-entry of the matrix $\mathbf{P}(t)$ represents the transition probability $P_{(Y,\nu)}(X,\pi;t)$.

For fixed  $X$ and $Y$,  we denote by $\mathbf{P}_Y(X;t)$ the $N^n \times N^n$ submatrix  of $\mathbf{P}(t)$ formed by restricting to all rows labelled $(X,\pi)$ and all columns labelled $(Y,\nu)$, where $\pi,\nu$ range over all  words of length $n$ with letters in $\{1,\dots, N\}$. Hence, the row and column of the matrix $\mathbf{P}_Y(X;t)$ are indexed by $\pi$ and $\nu$, respectively. Unless otherwise specified, we assume throughout this paper that any $N^n \times N^n$ matrix is indexed lexicographically, from $11\cdots1$ to $NN\cdots N$ in both rows and columns.

For notational convenience, we often abbreviate $\mathbf{P}_{Y}(X;t)$ as $\mathbf{P}(X;t)$, and likewise write $P_{\nu}(X,\pi;t)$ or $P(X,\pi;t)$ in place of $P_{(Y,\nu)}(X,\pi;t)$ when the initial state $(Y,\nu)$ is understood from context.

For an $N^n \times N^n$ matrix $\mathbf{P}(X;t)$, we define its time derivative $\frac{d}{dt}\mathbf{P}(X;t)$ as the matrix whose  $(\pi,\nu)$-entry  is given by $\frac{d}{dt}P_{\nu}(X,\pi;t)$.

Throughout this paper, $\otimes$ denotes the tensor (Kronecker) product of matrices.

\subsection{Natural extension to the $N$-species case}\label{224pm81}
\begin{definition}\label{733pm731}
Let $\mathbf{b}$ be a $4 \times 4$ matrix whose rows and column are labelled by $11,12,21,22$ representing the two-particle interaction at a common site between particles of species $1$ and $2$. The natural extension of $\mathbf{b}$ to the $N$-species case refers to the interpretation of these labels as $ii,ij,ji,jj$ for any pair $(i,j)$ with  $1\leq i<j \leq N$, such that   $\mathbf{b}$ describes the two-particle interaction at a common site between particles of species $i$ and  $j$.
\end{definition}
Not all matrices $\mathbf{b}^{(l)}$ in Appendix \ref{627pm729} admit a consistent natural extension to the $N$-species case.  For example, consider the matrix $\mathbf{b}^{(6)}$:
\begin{equation}\label{Bmatrix52133-728}
\mathbf{b}^{(6)}=  \kbordermatrix{
    & 11 & 12 & 21 & 22  \\
    11 & 0 & 0 & 0 & 0  \\
    12 & 0 & 0 & 0 & 0  \\
    21 & 0 & 1 & 1 & 0  \\
    22 & 1 & 0 & 0 & 1
  }.
\end{equation}
This matrix describes the interaction between particles of species $1$ and $2$. Now suppose we attempt to extent it to describe the interaction between particles of species $2$ and $3$. In that case, the corresponding matrix would be
\begin{equation}\label{Bmatrix934pm728}
 \kbordermatrix{
    & 22 & 23 & 32 & 33  \\
    22 & 0 & 0 & 0 & 0  \\
    23 & 0 & 0 & 0 & 0  \\
    32 & 0 & 1 & 1 & 0  \\
    33 & 1 & 0 & 0 & 1
  }.
\end{equation}
However, this leads to an inconsistency:  the $(22,22)$-entry of the original matrix (\ref{Bmatrix52133-728}) is $1$, while the  $(22,22)$-entry in (\ref{Bmatrix934pm728}) is $0$. Hence, $\mathbf{b}^{(6)}$ cannot be extended in a consistent manner under the natural extension. Consequently, only the matrices $\mathbf{b}^{(l)}$ with $l=1,2,3,4,5,11,13$ in Appendix \ref{627pm729} admit a consistent natural extension to the $N$-species case.

If a two-particle interaction matrix $\mathbf{b}$ of size $4 \times 4$ can be naturally extended to the $N$-species case, we define an $N^2 \times N^2$ matrix $\mathbf{B} = (b_{\pi,\nu})$,  where rows and columns are labelled by ordered pairs $\pi=\pi_1\pi_2$ and $\nu=\nu_1\nu_2$, respectively, with $\pi_1,\pi_2,\nu_1,\nu_2 \in \{1,\dots, N\}$, ordered   lexicographically from $11$ to $NN$.  Each entry $b_{\pi,\nu}$ is defined to be equal to the corresponding entry of $\mathbf{b}$ if both $\pi$ and $\nu$ belong to $\{ii,ij,ji,jj\}$ for some fixed $i<j$, and zero, otherwise.
\subsection{Tensor product}\label{1059pm729}
Let $\mathbf{M} = (a_{i_1,j_1})$ be an $m_1 \times n_1$ matrix with $1\leq i_1 \leq m_1, 1\leq j_1 \leq n_1$ and $\mathbf{M}' = (b_{i_2,j_2})$ be an $m_2 \times n_2$ matrix with $1\leq i_2 \leq m_2, 1\leq j_2 \leq n_2$. Then, the tensor product, $\mathbf{M} \otimes \mathbf{M}'$, is an $m_1m_2 \times n_1n_2$ matrix whose rows are labeled by pairs $i_1i_2$ where $i_1=1,\dots,m_1$ and $i_2 = 1,\dots, m_2$ and columns are labeled by pairs $j_1j_2$ where $j_1=1,\dots,n_1$ and $j_2 = 1,\dots, n_2$. The entries of $\mathbf{M} \otimes \mathbf{M}'$ are given by
\begin{equation*}
(\mathbf{M} \otimes \mathbf{M}')_{i_1i_2,j_1j_2} = a_{i_1,j_1}b_{i_2,j_2}.
\end{equation*}
We will frequently use the notation $\mathbf{A}^{\otimes n}$ to denote the $n$-fold tensor product
\begin{equation*}
\underbrace{\mathbf{A}~\otimes~ \cdots ~\otimes ~\mathbf{A}}_{\textrm{$n$ times }}.
\end{equation*}

Let $\mathbf{I}$ denote the $N \times N$ identity matrix, and let $\mathbf{B}$ be the $N^2 \times N^2$ matrix describing two-particle interactions in the $N$-species case, as defined in the previous section. Then, the  matrix $\mathbf{B} \otimes \mathbf{I}$, of size $N^3 \times N^3$,  describes the interaction between the first and second particles, while leaving the third particle unchanged. The rows and columns of $\mathbf{B} \otimes \mathbf{I}$ are indexed  with triples $\pi=\pi_1\pi_2\pi_3$ and $\nu=\nu_1\nu_2\nu_3$  where each $\pi_i,\nu_i \in \{1,\cdots, N\}$, ordered lexicographically from $111$ to $NNN$. The entries are given by
\begin{equation*}
(\mathbf{B} \otimes \mathbf{I})_{\pi_1\pi_2\pi_3,\nu_1\nu_2\nu_3} =
\begin{cases}
(\mathbf{B})_{\pi_1\pi_2,\nu_1\nu_2}&~\textrm{if $\pi_3 = \nu_3$}\\
0&~\textrm{otherwise.}
\end{cases}
\end{equation*}
Similarly,  the matrix $\mathbf{I} \otimes \mathbf{B}$ describes the interaction between  the second and third particles, leaving the first particle unchanged. Its entries are given by
\begin{equation*}
(\mathbf{I} \otimes \mathbf{B})_{\pi_1\pi_2\pi_3,\nu_1\nu_2\nu_3} =
\begin{cases}
(\mathbf{B})_{\pi_2\pi_3,\nu_2\nu_3}&~\textrm{if $\pi_1 = \nu_1$}\\
0&~\textrm{otherwise.}
\end{cases}
\end{equation*}
More generally, in an $n$-particle system, the $N^n \times N^n$ matrix
\begin{equation}
\label{501pm727}
\begin{aligned}
\mathbf{B}_i:=\underbrace{\mathbf{I}~ \otimes~ \cdots ~\otimes ~\mathbf{I}}_{\textrm{$(i-1)$ times}} ~\otimes~ \mathbf{B}~\otimes~ \underbrace{\mathbf{I}~ \otimes~ \cdots~ \otimes~ \mathbf{I}}_{\textrm{$(n-i-1)$ times}}
\end{aligned}
\end{equation}
describes the interaction of the $i$th  and $(i+1)$th particles, leaving all other particles remaining unaffected.
\section{Integrability}\label{main}
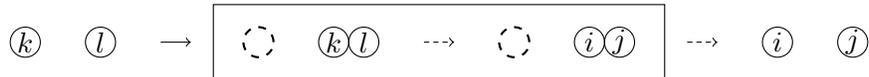
\begin{figure}[H]
  \centering
  \begin{tikzpicture}

    \draw[draw=black] (2.5,0.5) rectangle (8.5,-0.5);

    \node[circle, draw, minimum size=0.4cm, inner sep = 0] (left1) at (0,0) {$k$};
    \node[circle, draw, minimum size=0.4cm, inner sep = 0] (left2) at (1,0) {$l$};

    \node[circle, draw, dashed, thick, minimum size=0.4cm, inner sep = 0] (mid1) at (3.1,0) {};
    \node[circle, draw, minimum size=0.4cm, inner sep = 0] (mid2) at (4.1,0) {$k$};
    \node[circle, draw, minimum size=0.4cm, inner sep = 0] at (4.5,0) {$l$};

    \node[circle, draw, dashed, thick, minimum size=0.4cm, inner sep = 0] at (6.5,0) {};
    \node[circle, draw, minimum size=0.4cm, inner sep = 0] at (7.5,0) {$i$};
    \node[circle, draw, minimum size=0.4cm, inner sep = 0] at (7.9,0) {$j$};

    \node[circle, draw, minimum size=0.4cm, inner sep = 0] at (10.0,0) {$i$};
    \node[circle, draw, minimum size=0.4cm, inner sep = 0] at (11.0,0) {$j$};

    \draw[->] (1.8,0) -- (2.2,0);
    \draw[->, dash pattern=on 2pt off 1.5pt] (5.3,0) -- (5.7,0);
    \draw[->, dash pattern=on 2pt off 1.5pt] (8.8,0) -- (9.2,0);
  \end{tikzpicture}
  \caption{The particle $i$ in the box jumps backward. }
  \label{fig:fig2}
\end{figure}

Let us consider an $n$-particle system in which the two-particle interaction is governed by the natural extension  of one of the matrices $\mathbf{b}^{(l)}$  for $l=1,2,3,4,5,11,13$ in Appendix \ref{627pm729}. We assume that after a two-particle interaction occurs at a common site,  the particle located on the left immediately jumps backward as illustrated  in Figure \ref{fig:fig2}.
\subsection{Master equations}\label{subsectionA}
The form of the master equation depends on the relative positions of the particles. Following a standard approach \cite{Ali-Ahmadi,Lee-2020,Lee-2024,Schutz-1997}, all possible configurations can be treated within a unified framework: when particles are well separated, the evolution is governed by a simple non–interaction equation, while the effects of interactions are fully captured by boundary conditions imposed when two or more particles occupy adjacent sites. In this way, the variety of master equations reduces to a single non–interaction equation together with boundary conditions that encode the multi–particle dynamics, which in turn decompose into successive two–particle interactions.
\subsubsection{Two-particle systems}

Two-particle systems serve as the building blocks for the general $n$-particle dynamics.
In particular, the interaction rules of the $n$-particle system are obtained by combining two-particle interactions through tensor products of the corresponding matrices.
Thus, once the two-particle case is understood, the structure of the general system follows naturally from the tensor product construction.

Recall that $\mathbf{B}$ denotes the $N^2 \times N^2$ matrix as defined in Section \ref{224pm81}.
The master equations of the transition probabilities $P_{(Y,\nu)}(X,\pi;t)$ where $\pi=\pi_1\pi_2$ and $\nu=\nu_1\nu_2$ with $\pi_1,\pi_2,\nu_1,\nu_2 \in \{1,\dots, N\}$ form a system of equations encoded in a matrix of size $N^2 \times N^2$, depending on the configuration $X=(x_1,x_2)$, as follows:
\begin{itemize}
  \item [(\rmnum{1})] if $X=(x_1,x_2)$ with $x_1< x_2-1$, then
  \begin{equation}\label{420845pm726}
\begin{aligned}
\frac{d}{dt}\mathbf{P}(x_1,x_2;t) =\,&  \mathbf{P}(x_1-1,x_2;t) + \mathbf{P}(x_1,x_2-1;t)  - 2\mathbf{P}(x_1,x_2;t),
\end{aligned}
\end{equation}
  \item [(\rmnum{2})] if $X=(x,x+1)$, then
  \begin{equation}\label{4197088pm726}
\begin{aligned}
\frac{d}{dt}\mathbf{P}(x,x+1;t) =\,&  \mathbf{P}(x-1,x+1;t) + \mathbf{B}\mathbf{P}(x,x+1;t)  - 2\mathbf{P}(x,x+1;t).
\end{aligned}
\end{equation}
\end{itemize}

Now, let $\mathbf{U}(x_1,x_2;t) = (u_{\pi,\nu})$ be an $N^2 \times N^2$ matrix-valued function where each entry $u_{\pi,\nu}$ is a function  $U_{\nu}(x_1,x_2,\pi;t)$ defined for all $(x_1,x_2) \in \mathbb{Z}^2$ and $t \in [0,\infty)$. Suppose that $\mathbf{U}(x_1,x_2;t)$ satisfies
\begin{equation*}
\frac{d}{dt}\mathbf{U}(x_1,x_2;t) =  \mathbf{U}(x_1-1,x_2;t) + \mathbf{U}(x_1,x_2-1;t)  - 2\mathbf{U}(x_1,x_2;t)
\end{equation*}
for all $(x_1,x_2) \in \mathbb{Z}^2$, and the boundary condition
\begin{equation}\label{4201000pm726}
\mathbf{U}(x,x;t)= \mathbf{B}\mathbf{U}(x,x+1;t)
\end{equation}
for all $x \in \mathbb{Z}$. Then, $\mathbf{U}(x_1,x_2;t)$ satisfies the master equation (\ref{420845pm726}) and (\ref{4197088pm726}) for $X=(x_1,x_2)$ with $x_1< x_2-1$ and $X=(x,x+1)$, respectively.
\subsubsection{General $n$-particle systems}
We first consider the case for $n=3$, followed by the general case. Let us recall the interpretation of the matrices $\mathbf{B}\otimes \mathbf{I}$ and $\mathbf{I}\otimes\mathbf{B}$, as discussed in Section \ref{1059pm729}. Then, we observe that the master equations for $P_{\nu}(x_1,x_2,x_3,\pi;t)$ can be expressed in matrix forms of size $N^3 \times N^3$, depending on the configuration $X= (x_1,x_2,x_3)$ as follows:
\begin{itemize}
  \item [(\rmnum{1})] if $X = (x_1,x_2,x_3)$ with $x_i< x_{i+1}-1,\,(i=1,2)$, then
  \begin{equation}\label{422-628-pm}
\begin{aligned}
 \frac{d}{dt}\mathbf{P}(x_1,x_2,x_3;t) =\,&  \mathbf{P}(x_1-1,x_2,x_3;t) + \mathbf{P}(x_1,x_2-1,x_3;t)\\[4pt]
 &~+ \mathbf{P}(x_1,x_2,x_3-1;t)- 3\mathbf{P}(x_1,x_2,x_3;t).
\end{aligned}
\end{equation}
  \item [(\rmnum{2})] if   $X = (x,x+1, x_3)$ with $x+1< x_3-1$, then
  \begin{equation}\label{42222-633-pm}
\begin{aligned}
 \frac{d}{dt}\mathbf{P}(x,x+1, x_3;t) =\, & \mathbf{P}(x-1,x+1,x_3;t) +   \mathbf{P}(x,x+1,x_3-1;t)  \\[4pt]
                                      &+(\mathbf{B}\otimes \mathbf{I})\mathbf{P}(x,x+1,x_3;t)- 3\mathbf{P}(x,x+1,x_3;t)
\end{aligned}
\end{equation}
  \item [(\rmnum{3})] if $X = (x_1,x,x+1)$ with $x_1<x-1$, then
   \begin{equation}\label{423-425-pm}
\begin{aligned}
\frac{d}{dt}\mathbf{P}(x_1,x,x+1;t) =\,&  \mathbf{P}(x_1-1,x,x+1;t) +  \mathbf{P}(x_1,x-1,x+1;t)\\[4pt]
                                    &+ (\mathbf{I}\otimes\mathbf{B})\mathbf{P}(x_1,x,x+1;t)- 3\mathbf{P}(x_1,x,x+1;t)
\end{aligned}
\end{equation}
  \item [(\rmnum{4})] if $X= (x,x+1,x+2)$, then
   \begin{equation}\label{429-633-pm}
\begin{aligned}
 \frac{d}{dt}\mathbf{P}(x,x+1, x+2;t) =\,&  \mathbf{P}(x-1,x+1,x+2;t) +(\mathbf{B}\otimes \mathbf{I}) \mathbf{P}(x,x+1,x+2;t) \\[4pt]
                                    & + (\mathbf{I}\otimes\mathbf{B})\mathbf{P}(x,x+1,x+2;t)- 3\mathbf{P}(x,x+1,x_3;t).
\end{aligned}
\end{equation}
\end{itemize}

Let $\mathbf{U}(x_1,x_2,x_3;t) = (u_{\pi,\nu})$ be an $N^3 \times N^3$ matrix whose entries are given by functions $u_{\pi,\nu}=U_{\nu}(x_1,x_2,x_3,\pi;t)$ defined for all $(x_1,x_2,x_3) \in \mathbb{Z}^3$ and $t \in [0,\infty)$. Suppose that $\mathbf{U}(x_1,x_2,x_3;t)$ satisfies
 \begin{equation}\label{555pm67}
\begin{aligned}
 \frac{d}{dt}\mathbf{U}(x_1,x_2,x_3;t) =\,&  \mathbf{U}(x_1-1,x_2,x_3;t) + \mathbf{U}(x_1,x_2-1,x_3;t)\\[4pt]
 &~+ \mathbf{U}(x_1,x_2,x_3-1;t)- 3\mathbf{U}(x_1,x_2,x_3;t)
\end{aligned}
\end{equation}
and the boundary conditions
\begin{equation}\label{4201000pm522}
\begin{aligned}
& \mathbf{U}(x,x,x';t)= (\mathbf{B}\otimes \mathbf{I})\mathbf{U}(x,x+1,x';t), \\
& \mathbf{U}(x',x,x;t)= ( \mathbf{I}\otimes\mathbf{B})\mathbf{U}(x',x,x+1;t)
\end{aligned}
\end{equation}
for all $x,x' \in \mathbb{Z}$. Then, $\mathbf{U}(x_1,x_2,x_3;t)$ satisfies the master equations (\ref{422-628-pm}),(\ref{42222-633-pm}),(\ref{423-425-pm}) and (\ref{429-633-pm}), respectively, for each corresponding configuration $X=(x_1,x_2,x_3)$.

This framework for $n=2$ and $n=3$ naturally extends to the general $n$-particle case for $n\geq 2$ as in the previous works on the TASEP or the ASEP \cite{Schutz-1997,Tracy-Widom-2008}.

Let $\mathbf{U}(x_1,\dots,x_n;t) = (u_{\pi,\nu})$ denote an $N^n \times N^n$ matrix whose entries are given by  $u_{\pi,\nu}=U_{\nu}(x_1,\dots,x_n,\pi;t)$, defined for all $(x_1,\dots,x_n) \in \mathbb{Z}^n$ and $t \in [0,\infty)$. Suppose that  $\mathbf{U}(x_1,\dots,x_n;t)$ satisfies
\begin{equation}\label{329pm69}
\begin{aligned}
\frac{d}{dt}\mathbf{U}(x_1,\dots,x_n;t) =\,& \mathbf{U}(x_1-1,x_2,\dots,x_n;t) + \mathbf{U}(x_1,x_2-1,x_3,\dots,x_n;t)\\[4pt]
& \hspace{0.5cm}+ \cdots + \mathbf{U}(x_1,\dots,x_{n-1},x_n-1;t) - n\mathbf{U}(x_1,\dots,x_n;t),
\end{aligned}
\end{equation}
and  satisfies the boundary condition, for each $i=1,\dots, n-1$,
\begin{equation}\label{559pm610}
\mathbf{U}(x_1,\dots,x_{i-1},x_i,x_i,x_{i+2},\dots, x_n;t) = \mathbf{B}_i\mathbf{U}(x_1,\dots,x_{i-1},x_i,x_i+1,x_{i+2},\dots, x_n;t)
\end{equation}
where $\mathbf{B}_i$ is defined in (\ref{501pm727}). Then, $\mathbf{U}(x_1,\dots,x_n;t)$ satisfies the master equation that $\mathbf{P}(X;t)$ satisfies  for each configuration $X = (x_1,\dots,x_n)$ with $x_1< \cdots <x_n$.
\subsection{Yang-Baxter integrability}\label{YB}
The Bethe ansatz solution for the $(\pi,\nu)$-entry of the matrix equation (\ref{329pm69})  is given by
\begin{equation}\label{347pm69}
U_{\nu}(x_1,\dots,x_n,\pi;t) = \sum_{\sigma \in \mathcal{S}_n} A_{\sigma,(\pi,\nu)}\prod_{i=1}^n\xi_{\sigma(i)}^{x_i} e^{\varepsilon(\xi_i)t}
\end{equation}
for any collection of nonzero complex parameters $\xi_1,\cdots,\xi_n$. Here,  $\mathcal{S}_n$ denotes the symmetric group on $n$ elements, $A_{\sigma,(\pi,\nu)}$ are constants independent of the variables $x_i$ and $t$, and
\begin{equation}\label{234pm618}
\varepsilon(\xi_i) = \frac{1}{\xi_i} -1.
\end{equation}
Denoting by $\mathbf{A}_{\sigma}$ the matrix whose $(\pi,\nu)$-entry is $A_{\sigma,(\pi,\nu)}$, we may express the full matrix-valued solution to equation (\ref{329pm69}) in compact form as
\begin{equation}\label{600pm610}
\mathbf{U}(x_1,\dots,x_n;t) = \sum_{\sigma \in \mathcal{S}_n} \mathbf{A}_{\sigma}\prod_{i=1}^n\xi_{\sigma(i)}^{x_i} e^{\varepsilon(\xi_i)t}.
\end{equation}
This ansatz also must   satisfy the boundary conditions (\ref{559pm610}) for each $i=1,\dots, n-1$.
\begin{definition}\label{1204am730}
Given an $N^2 \times N^2$  two-particle interaction matrix $\mathbf{B}$,   we define an $N^2 \times N^2$ matrix
\begin{equation}\label{623pm723}
\mathbf{R}_{\beta\alpha} = -(\mathbf{I}\otimes \mathbf{I} - \xi_{\alpha}\mathbf{B})^{-1}(\mathbf{I}\otimes \mathbf{I} - \xi_{\beta}\mathbf{B}),
\end{equation}
where $\mathbf{I}$ is the $N \times N$ identity matrix and define an $N^n \times N^n$ matrix
\begin{equation}\label{803pm68}
\mathbf{T}_{i,\beta\alpha} := \underbrace{\mathbf{I} ~\otimes~\cdots  ~\otimes~ \mathbf{I}}_{\textrm{$(i-1)$ times}} ~\otimes ~ \mathbf{R}_{\beta\alpha}  ~\otimes~ \underbrace{\mathbf{I}~\otimes ~\cdots~ \otimes ~ \mathbf{I}}_{\textrm{$(n-i-1)$ times}}
\end{equation}
each $i = 1,\dots,n-1$.
\end{definition}
\begin{remark}
If $\mathbf{B}$ is defined by the specific form (\ref{623pm720}), for example, then the corresponding matrices $\mathbf{R}_{\beta\alpha}$ and $\mathbf{T}_{i,\beta\alpha}$ reduce to those appearing in the mTASEP developed  in \cite{Lee-2020}.
\end{remark}
Let $T_i\in \mathcal{S}_n$ for $i=1,\dots, n-1$ denote the adjacent transposition that swaps the $i$th and the $(i+1)$th elements, leaving all other elements unchanged. It is well known that the symmetric group $\mathcal{S}_n$ is generated by these adjacent transpositions. Consequently,  any permutation $\sigma \in \mathcal{S}_n$ can be written as a product of adjacent permutations:
\begin{equation}\label{157pm610}
\sigma = T_{i_k}\cdots T_{i_1}
\end{equation}
for some sequence $i_1,\dots, i_k$ with $i_j \in \{1,\dots, n-1\}$. The representation (\ref{157pm610}) is not unique, but any two such expressions for the same permutation are related to a finite sequence of braid relations,
\begin{equation}\label{636pm727}
\begin{aligned}
& T_iT_i = ~\textrm{the identity},\\
& T_iT_j = T_jT_i~~\textrm{if $|i-j|\geq 2$,}\\
& T_iT_{i+1}T_i = T_{i+1}T_iT_{i+1},
\end{aligned}
\end{equation}
which are defining relations of the symmetric group $\mathcal{S}_n$.

Let $\sigma = T_{i_k}\cdots T_{i_1}$ be a given decomposition of $\sigma$ into adjacent transpositions. For each $j = 1,\dots, k$,  let $(\beta_j,\alpha_j)$ denote the pair of elements swapped by $T_{i_j}$ when acting on the current configuration of the evolving permutation, that is,
\begin{equation*}
T_{i_j}(\,\cdots\,\alpha_j\,\beta_j\,\cdots\,) = (\,\cdots\,\beta_j\,\alpha_j\,\cdots\,).
\end{equation*}
We then define
\begin{equation}\label{232pm618}
\mathbf{A}_{\sigma}:= \mathbf{T}_{i_k,\beta_k\alpha_k} \times \cdots \times  \mathbf{T}_{i_1,\beta_1\alpha_1}.
\end{equation}
The well-definedness of $\mathbf{A}_{\sigma}$, that is, its independence of the choice of reduced decomposition of $\sigma$, must be verified using the braid relations (\ref{636pm727}).
\begin{proposition}
If (\ref{232pm618}) is well-defined, then
 $\mathbf{U}(x_1,\dots,x_n;t)$ in  (\ref{600pm610}) satisfies the boundary conditions (\ref{559pm610}) for all $i=1,\dots, n-1$.
\end{proposition}
\begin{proof}
We begin by noting that for any adjacent transposition $T_i$, the relation
\begin{equation}
\mathbf{A}_{T_i\sigma} = \mathbf{T}_{i,\sigma(i+1)\sigma(i)}\mathbf{A}_{\sigma}
\end{equation}
holds by the definition of $\mathbf{A}_{\sigma}$. Using the standard identities of tensor products,
\begin{equation*}
(A \otimes B)(C \otimes D) = AC \otimes BD~~\textrm{and}~~(A\otimes B)^{-1} = A^{-1} \otimes B^{-1}
\end{equation*}
which hold for matrices $A,B,C,D$ of appropriate dimensions, we obtain the following expression:
\begin{equation*}
\begin{aligned}
\mathbf{A}_{T_i\sigma} =~&\mathbf{T}_{i,\sigma(i+1)\sigma(i)}\mathbf{A}_{\sigma} =\big(\mathbf{I}^{\otimes (i-1)}~\otimes ~ \mathbf{R}_{\sigma(i+1)\sigma(i)} ~\otimes ~\mathbf{I}^{\otimes (n-i-1)}\big)\mathbf{A}_{\sigma}\\[4pt]
=~& -(\mathbf{I}^{\otimes n} - \mathbf{B}_i\xi_{\sigma(i)})^{-1}(\mathbf{I}^{\otimes n} - \mathbf{B}_i\xi_{\sigma(i+1)})\mathbf{A}_{\sigma}
\end{aligned}
\end{equation*}
where $\mathbf{B}_i$ is defined in (\ref{501pm727}). Rewriting this equation, we obtain that
\begin{equation*}
 \big(\mathbf{I}^{\otimes n} - \mathbf{B}_i\xi_{\sigma(i)}\big)\mathbf{A}_{T_i\sigma} + \big(\mathbf{I}^{\otimes n} - \mathbf{B}_i\xi_{\sigma(i+1)}\big)\mathbf{A}_{\sigma} = \mathbf{0},
\end{equation*}
for each $\sigma \in \mathcal{S}_n$. Summing both sides over $\sigma  \in \mathcal{A}_n$ where $\mathcal{A}_n$ denotes the alternating group, we obtain
\begin{equation*}
\begin{aligned}
\mathbf{0} = ~&\sum_{\sigma \in \mathcal{A}_n}\bigg(\big(\mathbf{I}^{\otimes n} - \mathbf{B}_i\xi_{\sigma(i)}\big)\mathbf{A}_{T_i\sigma} + \big(\mathbf{I}^{\otimes n} - \mathbf{B}_i\xi_{\sigma(i+1)}\big)\mathbf{A}_{\sigma}\bigg) \\
=~& \sum_{\sigma \in \mathcal{S}_n}\big(\mathbf{I}^{\otimes n} - \mathbf{B}_i\xi_{\sigma(i+1)}\big)\mathbf{A}_{\sigma},
\end{aligned}
\end{equation*}
where in the second equality we used the fact that the map $\sigma \longmapsto T_i\sigma$ is a bijection between $\mathcal{A}_{\sigma}$ and its complement in $\mathcal{S}_n$, thus covering the full symmetric group in the sum.  This identity is precisely equivalent to the equation obtained by substituting $\mathbf{U}(x_1,\dots,x_n;t)$ in (\ref{600pm610}) into the boundary condition (\ref{559pm610}).
\end{proof}
In order to show the well-definedness of (\ref{232pm618}), it suffices to show that the matrices $\mathbf{T}_{i,\beta\alpha}$ satisfy the analogues of the defining relations of the symmetric group $\mathcal{S}_n$ when $\sigma$ is represented by adjacent transpositions. The relations corresponding to (\ref{636pm727}) are:
\begin{equation}\label{603pm612}
\begin{aligned}
(a)~& \mathbf{T}_{i,\alpha\beta}\mathbf{T}_{i,\beta\alpha} = ~\textrm{the identity matrix},\\
(b)~& \mathbf{T}_{i,\beta\alpha}\mathbf{T}_{j,\gamma\delta} = \mathbf{T}_{j,\gamma\delta}\mathbf{T}_{i,\beta\alpha}~~~~\textrm{if}~|i-j|\geq 2, \\
(c)~&\mathbf{T}_{i,\gamma\beta}\mathbf{T}_{i+1,\gamma\alpha}\mathbf{T}_{i,\beta\alpha} = \mathbf{T}_{i+1,\beta\alpha}\mathbf{T}_{i,\gamma\alpha}\mathbf{T}_{i+1,\gamma\beta}.
\end{aligned}
\end{equation}
\begin{proposition}\label{1217pm728}
Let  $\mathbf{B}$ be the matrix defined by  $\mathbf{b}^{(l)}$ for $l=1,2,3,4,5,11,13$  in Appendix \ref{627pm729}.  Then, the matrices  $\mathbf{T}_{i,\beta\alpha}$  in Definition \ref{1204am730}  satisfy the relations $(a),(b),$ and $(c)$ in (\ref{603pm612}).
\end{proposition}
\begin{proof}
To verify relation $(a)$, it suffices to consider the case $n=2$.  In this case, it is straightforward to observe that
\begin{equation}\label{635pm612}
\begin{aligned}
\mathbf{R}_{\alpha\beta}\mathbf{R}_{\beta\alpha} = & ~(\mathbf{I}\otimes \mathbf{I} - \xi_{\beta}\mathbf{B})^{-1}(\mathbf{I}\otimes \mathbf{I} - \xi_{\alpha}\mathbf{B}) (\mathbf{I}\otimes \mathbf{I} - \xi_{\alpha}\mathbf{B})^{-1}(\mathbf{I}\otimes \mathbf{I} - \xi_{\beta}\mathbf{B})  \\
                                                   =&  ~\textrm{the  identity matrix}.
\end{aligned}
\end{equation}
Relation $(b)$ follows immediately from the tensor product structure of $\mathbf{T}_{i,\beta\alpha}$.
For relation $(c)$, it suffices to verify it in the minimal nontrivial case $N=n=3$, which corresponds to the following identity between $27 \times 27$ matrices:
\begin{equation}\label{636pm612}
(\mathbf{R}_{\gamma\beta} \otimes \mathbf{I})(\mathbf{I} \otimes \mathbf{R}_{\gamma\alpha})(\mathbf{R}_{\beta\alpha} \otimes \mathbf{I}) = (\mathbf{I} \otimes \mathbf{R}_{\beta\alpha})(\mathbf{R}_{\gamma\alpha} \otimes \mathbf{I})(\mathbf{I} \otimes \mathbf{R}_{\gamma\beta}).
\end{equation}
Direct matrix multiplication confirms that relation $(c)$ holds for $l=1,2,3,4,5,11,13$.
\end{proof}

\begin{remark}
Relation~(c) above is the celebrated Yang-Baxter equation. It is a fundamental consistency condition originating in exactly solvable models of statistical mechanics.
In probabilistic terms, it ensures that the outcome of successive two-particle interactions is independent of the order in which they occur, and it serves as the algebraic foundation for integrability in many interacting particle systems.
\end{remark}

\subsection{Generalizations}\label{202pm81}
We have verified that the matrices $\mathbf{b}^{(l)}$ for $l=1,2,3,4,5,11,13$  in Appendix \ref{627pm729} can  be naturally extended to integrable $N$-species models. This leads to a natural question: does a convex linear combination of these matrices also result in an integrable $N$-species model? To investigate this, we examined with the help of a symbolic computer software whether the relations in (\ref{603pm612}), in particular the relation $(c)$, the Yang-Baxter equation, are preserved under convex linear combinations of $\mathbf{b}^{(i)}$ and  $\mathbf{b}^{(j)}$  for each pair $(i,j)$ with $i,j=1,2,3,4,5,11,13$. We found that only the  following convex linear combinations $\mathbf{b}^{(i)}$ and  $\mathbf{b}^{(j)}$ can be naturally extended to $N$-species integrable models:
\begin{itemize}
  \item [(i)]  $\mathbf{b}^{(1)}$ and  $\mathbf{b}^{(j)}$ for each $j=1,2,3,4,5,11,13.$
  \item [(ii)] $\mathbf{b}^{(3)}$ and  $\mathbf{b}^{(j)}$ for $j=4,5$.
  \item [(iii)] $\mathbf{b}^{(4)}$ and  $\mathbf{b}^{(j)}$ for $j=5,11$.
\end{itemize}
In case (i), one can also verify algebraically that relation $(c)$ in (\ref{603pm612})is satisfied. Specifically, suppose that $\mathbf{R}_{\beta\alpha}$ defined using the $N$-species extension $\mathbf{B}$ of $\mathbf{b}^{(j)}$, satisfies the Yang-Baxter equation (\ref{636pm612}). Then, consider  a convex linear combination $a\mathbf{b}^{(1)} +b \mathbf{b}^{(j)}$ with $a+b=1$. Its $N$-species extension corresponds to the matrix  $a\mathbf{I}\otimes \mathbf{I} + b\mathbf{B}$, and the associated  scattering matrix becomes
\begin{equation}\label{1034am730}
 -\frac{1-a\xi_{\beta}}{1-a\xi_{\alpha}}\bigg(\mathbf{I}\otimes \mathbf{I} - \frac{b\xi_{\alpha}}{1-a\xi_{\alpha}}\mathbf{B}\bigg)^{-1}\bigg(\mathbf{I}\otimes \mathbf{I} - \frac{b\xi_{\beta}}{1-a\xi_{\beta}}\mathbf{B}\bigg).
\end{equation}
This expression results from a simple rescaling and reparameterization of the original formula (\ref{623pm723}). Since these operations preserve the algebraic structure of the Yang–Baxter equation, the integrability of the model is maintained.

 For all other pairs $(i,j)$ with $i\neq j$ and $i,j \in \{1,2,3,4,5,11,13\}$ not covered in cases (i),(ii), or (iii), the scattering matrix corresponding to the linear combination of $\mathbf{b}^{(i)}$  and $\mathbf{b}^{(j)}$  fails to satisfy the Yang-Baxter equation. These results underscore the delicate algebraic structure underlying integrable models and highlight specific families of convex linear combinations that preserve it.

The following result provides a broader generalization that encompasses most of the  linear combinations appearing in (i),(ii), and (iii). Consider the matrix
\begin{equation}\label{1058am730}
\left(
  \begin{array}{cccc}
    1 & 0 & 0 & 0 \\
    0 & \lambda & 0 & 0 \\
    0 & 0 & \lambda' & 0 \\
    0 & 1-\lambda & 1- \lambda'  & 1 \\
  \end{array}
\right)
\end{equation}
where $0\leq \lambda,\lambda'\leq 1$. This matrix includes the convex linear combinations of $\mathbf{b}^{(i)}$ and  $\mathbf{b}^{(j)}$ corresponding to the  pairs $(1,j)$ for $j=1,3,4,5,$ as well as $(3,4),(3,5)$ and $(4,5)$. While a convex linear combination $a\mathbf{b}^{(i)} + (1-a)\mathbf{b}^{(j)}$ involves a single parameter $a$,  the matrix (\ref{1058am730}) introduces two independent  parameters $\lambda$ and $\lambda'$, thus allowing greater flexibility and a richer class of interaction structures within the integrable framework.

The corresponding $N$-species model defined by the matrix (\ref{1058am730}) evolves according to the following dynamics:
\begin{quote}
Each particle of species $i$ at $x$ independently jumps to site $x+1$ after an exponential waiting time with rate $1$. If $x+1$ is occupied by a particle of species $j$, then $x+1$ temporarily accommodates both particles, with species $i$ occupying the left position and species $j$ the right position. If $i\neq j$, the following \textit{infection} rules apply:
\begin{itemize}
  \item [(a)] If $j<i$, then the particle of species $j$ is converted to  species $i$ with probability $1-\lambda'$, and remains unchanged with probability $\lambda'$.
  \item [(b)] If $j>i$, then the particle of species $i$ is converted to species $j$ with probability $1-\lambda$, and remains unchanged  with probability $\lambda$.
\end{itemize}
Immediately after this interaction, the left particle moves back to $x$.
\end{quote}

Let $\mathbf{R}_{\beta\alpha} = (R_{\pi_1\pi_2,\nu_1\nu_2})$ denote the $N^2 \times N^2$ scattering matrix  defined by the matrix (\ref{1058am730}). Then, the entries of $\mathbf{R}_{\beta\alpha}$ take the form
 \begin{equation*}
R_{\pi_1\pi_2,\nu_1\nu_2} =
\begin{cases}
S_{\beta\alpha}(1)&~\textrm{if $\pi_1\pi_2 =\nu_1\nu_2$ and $\nu_1 = \nu_2$,}\\
S_{\beta\alpha}(\lambda)&~\textrm{if $\pi_1\pi_2 =\nu_1\nu_2$ and $\nu_1 < \nu_2$,}\\
T_{\beta\alpha}(1-\lambda)&~\textrm{if $\pi_1\pi_2 =\nu_2\nu_2$ and $\nu_1 < \nu_2$,}\\
T_{\beta\alpha}(1-\lambda')&~\textrm{if $\pi_1\pi_2 =\nu_1\nu_1$ and $\nu_1 > \nu_2$,}\\
S_{\beta\alpha}(\lambda')&~\textrm{if $\pi_1\pi_2 =\nu_1\nu_2$ and $\nu_1 > \nu_2$,}\\
0&~\textrm{otherwise},
\end{cases}
\end{equation*}
where
\begin{equation*}
S_{\beta\alpha}(r) = -\frac{1-r\xi_{\beta}}{1-r\xi_{\alpha}}~~\textrm{and}~~T_{\beta\alpha}(r) = \frac{r(\xi_{\beta}-\xi_{\alpha})}{(1-\xi_{\alpha})(1-(1-r)\xi_{\alpha})}.
\end{equation*}
For instance, when $N=3$,
\\
\begin{equation*}
{\small
\mathbf{R}_{\beta\alpha} =  \kbordermatrix{
       & 11 & 12 & 13 & 21 & 22 & 23 & 31 & 32 & 33  \\[5pt]
   11  & S_{\beta\alpha}(1) & 0 & 0 & 0 & 0 & 0 & 0 & 0 & 0   \\[5pt]
   12  & 0 & S_{\beta\alpha}(\lambda) & 0 & 0 & 0 & 0 & 0 & 0 & 0 \\[5pt]
   13  & 0 & 0 & S_{\beta\alpha}(\lambda) & 0 & 0 & 0 & 0 & 0 & 0  \\[5pt]
   21  & 0 & 0 & 0 & S_{\beta\alpha}(\lambda') & 0 & 0 & 0 & 0 & 0  \\[5pt]
   22  & 0 & T_{\beta\alpha}(\mu) & 0 & T_{\beta\alpha}(\mu') & S_{\beta\alpha}(1) & 0 & 0 & 0 & 0  \\[5pt]
   23  & 0 & 0 & 0 & 0 & 0 & S_{\beta\alpha}(\lambda) & 0 & 0 & 0  \\[5pt]
   31  & 0 & 0 & 0 & 0 & 0 & 0 & S_{\beta\alpha}(\lambda') & 0 & 0 \\[5pt]
   32  & 0 & 0 & 0 & 0 & 0 & 0 & 0 & S_{\beta\alpha}(\lambda') & 0  \\[5pt]
   33  & 0 & 0 & T_{\beta\alpha}(\mu) & 0 & 0 & T_{\beta\alpha}(\mu) & T_{\beta\alpha}(\mu') & T_{\beta\alpha}(\mu') & S_{\beta\alpha}(1)
   }
}
\end{equation*}
\\
where $\mu = 1-\lambda$ and $\mu' = 1- \lambda'$.
It can be verified by using a symbolic computer software that the matrix $\mathbf{R}_{\beta\alpha}$ with $N=3$ which is the minimal case of the multi-species extending satisfies
\begin{equation*}
(\mathbf{R}_{\gamma\beta} \otimes \mathbf{I})(\mathbf{I} \otimes \mathbf{R}_{\gamma\alpha})(\mathbf{R}_{\beta\alpha} \otimes \mathbf{I}) = (\mathbf{I} \otimes \mathbf{R}_{\beta\alpha})(\mathbf{R}_{\gamma\alpha} \otimes \mathbf{I})(\mathbf{I} \otimes \mathbf{R}_{\gamma\beta}).
\end{equation*}
where $\mathbf{I}$ is the $3 \times 3$ identity matrix, thereby confirming   the integrability of the $N$-species model associated with the  matrix (\ref{1058am730}).

\begin{remark}
A simple way to view the interaction rules is as an infection process: when a higher–ranked particle meets a lower–ranked one, it “infects” the latter, so that the lower species changes into the higher one. In this sense, particles compete not only for space but also for type, with stronger species spreading at the expense of weaker ones. This provides an intuitive picture of the dynamics beyond the algebraic description.
\end{remark}

\subsection{Asymmetric extension to $N$-species cases}\label{452pm81}
We have established that the natural extension of $\mathbf{b}^{(l)}$ in Appendix \ref{627pm729} to the $N$-species case cannot be consistently defined for $l\neq 1,2,3,4,5,11,13$. This inconsistency arises from the presence of  certain nonzero entries such as the entries  $(ij,11)$ with $ij\neq 11$ or $(ij,22)$ with $ij\neq 22$  in the matrices $\mathbf{b}^{(l)}$ for these values of $l$. Such entries obstruct a coherent extension to $N$-species models as demonstrated in Section \ref{224pm81}. To address this,  we now propose an alternative extension scheme that accommodates a broader class of matrices.

We restrict our consideration to those matrices $\mathbf{b}^{(l)}$ whose fourth column is given by $(\,0~0~0~1\,)^t$. For such a matrix $\mathbf{b}^{(l)}$, we define a modified matrix $\mathbf{c}^{(l)}$  by replacing the first column with $(\,1~0~0~0\,)^t$. For example, given
\begin{equation*}
\mathbf{b}^{(14)} = \kbordermatrix{
    & 11 & 12 & 21 & 22  \\
    11 & 0 & 0 & 0 & 0  \\
    12 & 1 & 1 & 0 & 0  \\
    21 & 0 & 0 & 0 & 0  \\
    22 & 0 & 0 & 1 & 1
  },
\end{equation*}
the corresponding modified matrix becomes
\begin{equation*}
\mathbf{c}^{(14)} = \kbordermatrix{
    & ii & ij & ji & jj  \\
    ii & 1 & 0 & 0 & 0  \\
    ij & 0 & 1 & 0 & 0  \\
    ji & 0 & 0 & 0 & 0  \\
    jj & 0 & 0 & 1 & 1
  }.
\end{equation*}
We now consider an $N$-species model in which the interaction between two particles of species $1$ is  governed by $\mathbf{b}^{(l)}$, while all other pairwise interactions are governed by $\mathbf{c}^{(l)}$. For instance, for $\mathbf{b}^{(14)}$ with $N=3$,  the resulting $N^2 \times N^2$ matrix describing all two-particle interactions is given by
 \begin{equation*}
 \kbordermatrix{
       & 11 & 12 & 13 & 21 & 22 & 23 & 31& 32& 33\\
    11 & 0  & 0  & 0  & 0  & 0  & 0  & 0 & 0 & 0 \\
    12 & 1  & 1  & 0  & 0  & 0  & 0  & 0 & 0 & 0 \\
    13 & 0  & 0  & 1  & 0  & 0  & 0  & 0 & 0 & 0 \\
    21 & 0  & 0  & 0  & 0  & 0  & 0  & 0 & 0 & 0 \\
    22 & 0  & 0  & 0  & 1  & 1  & 0  & 0 & 0 & 0 \\
    23 & 0  & 0  & 0  & 0  & 0  & 1  & 0 & 0 & 0 \\
    31 & 0  & 0  & 0  & 0  & 0  & 0  & 0 & 0 & 0 \\
    32 & 0  & 0  & 0  & 0  & 0  & 0  & 0 & 0 & 0 \\
    33 & 0  & 0  & 0  & 0  & 0  & 0  & 1 & 1 & 1 \\
  }.
 \end{equation*}
Using a symbolic computation software, we examined whether the corresponding $N^2 \times N^2$ scattering matrices satisfy the Yang-Baxter equation for each $l=1,\dots, 28$ in order to determine the integrability of the associated models.  We found that integrability holds precisely when $\mathbf{c}^{(l)}= \mathbf{b}^{(k)}$ or $\mathbf{\overline{b}}^{(k)}$ for some $k \in  \{1,2,3,4,5,11,13\}$ where the notation $\mathbf{\overline{b}}^{(k)}$ was introduced in Remark \ref{926pm82}. This criterion is satisfied for $l=6,7,8,9,12,14,17,19$ as well as the trivial cases $l=1,2,3,4,5,11,13$. For all other values of $l$, the resulting model is not integrable.

\begin{remark}
We chose the vector $(\,1~0~0~0\,)^t$
 because it leads to a matrix that admits a natural extension to the multi-species case. This provided a good indication that the asymmetric modification was consistent, and indeed, upon testing, the scheme worked and preserved integrability.
\end{remark}

This asymmetric extension scheme is particularly interesting because it breaks the uniformity typically assumed in natural extensions, allowing one designated species (here, species 1) to retain a distinct interaction rule. Remarkably, despite this asymmetry, the resulting $N$-species models can still exhibit integrability. This suggests that integrability does not necessarily require complete symmetry among species and can persist under selective, structured deviations from uniform interactions. Our findings show that for certain choices of $\mathbf{b}^{(l)}$, such asymmetrically extended models yield scattering matrices that satisfy the Yang–Baxter equations, thereby confirming their integrability.

\begin{remark}
Beyond the natural extensions and generalizations discussed in Section \ref{202pm81}, as well as the asymmetric extensions possible for certain $\mathbf{b}^{(l)}$ matrices in this section, it remains unclear whether the remaining cases admit alternative extensions to the multi–particle setting. Exploring such possibilities could reveal further integrable structures, and it would be interesting to pursue this direction in future work.
\end{remark}

\section{Discussion}\label{discussion}
\subsection*{Models with the drop-push rule}
\begin{figure}[H]
\centering
\begin{tikzpicture}

    \draw[draw=black] (2.5,0.5) rectangle (8.5,-0.5);

    \node[circle, draw, minimum size=0.4cm, inner sep = 0] (left1) at (0,0) {$k$};
    \node[circle, draw, minimum size=0.4cm, inner sep = 0] (left2) at (1,0) {$l$};

    \node[circle, draw, dashed, thick, minimum size=0.4cm, inner sep = 0] (mid1) at (3.1,0) {};
    \node[circle, draw, minimum size=0.4cm, inner sep = 0] (mid2) at (4.1,0) {$k$};
    \node[circle, draw, minimum size=0.4cm, inner sep = 0] at (4.5,0) {$l$};

    \node[circle, draw, dashed, thick, minimum size=0.4cm, inner sep = 0] at (6.5,0) {};
    \node[circle, draw, minimum size=0.4cm, inner sep = 0] at (7.5,0) {$i$};
    \node[circle, draw, minimum size=0.4cm, inner sep = 0] at (7.9,0) {$j$};

    \node[circle, draw, dashed, thick, minimum size=0.4cm, inner sep = 0] at (10.0,0) {};
    \node[circle, draw, minimum size=0.4cm, inner sep = 0] at (11.0,0) {$i$};
    \node[circle, draw, minimum size=0.4cm, inner sep = 0] at (12.0,0) {$j$};

    \draw[->] (1.8,0) -- (2.2,0);
    \draw[->, dash pattern=on 2pt off 1.5pt] (5.3,0) -- (5.7,0);
    \draw[->, dash pattern=on 2pt off 1.5pt] (8.8,0) -- (9.2,0);
  \end{tikzpicture}

  \caption{Particle $j$ jumping forward after two-particle interaction at a common site}
  \label{fig:fig123}
\end{figure}
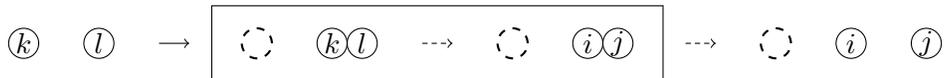

In the models discussed so far, the left particle at the common site jumps backward after the interaction (see Figure \ref{fig:fig2}). In contrast, we now consider a different dynamics where the right particle jumps forward after the interaction (see Figure \ref{fig:fig123}). In the single-species case, this is  known as the drop-push model \cite{Schutz-Ramaswamy-Barma}, and its asymptotic result and various generalizations have been studied \cite{Ali,Ali2,Borodin2}.
Switching from the backward-jumping rule (as in TASEP) to the forward-jumping rule (as in the drop-push model) does not affect the integrability of the system. In fact,  \cite{Lee-2024} demonstrates that the $N$-species model governed by the matrix $\mathbf{b}^{(2)}$ in Appendix \ref{627pm729}, when combined with the drop-push rule, remains integrable. In line with the discussion in \cite{Lee-2024}, if $\mathbf{B}$ is any of the integrable $N^2 \times N^2$ matrices describing two-particle interactions found in Section \ref{202pm81} or \ref{452pm81}, then the model defined by $\mathbf{B}$  and the drop-push rule is also integrable. The necessary modifications from the backward jumping case are minimal:

First, the boundary condition (\ref{559pm610}) is replaced by
\begin{equation}\label{559pm81}
\mathbf{U}(x_1,\dots,x_{i-1},x_i,x_i,x_{i+2},\dots, x_n;t) = \mathbf{B}_i\mathbf{U}(x_1,\dots,x_{i-1},x_i-1,x_i,x_{i+2},\dots, x_n;t).
\end{equation}
It was shown in \cite{Lee-2024} that   if  a function $\mathbf{U}(x_1,\dots,x_n;t)$ satisfies the evolution equation (\ref{329pm69}) along with the boundary conditions (\ref{559pm81}) for the case of $\mathbf{b}^{(2)}$, then  it satisfies the master equation that $\mathbf{P}_Y(X;t)$ satisfies  for each configuration $X = (x_1,\dots,x_n)$ with $x_1< \cdots <x_n$. Although this result was originally demonstrated for $\mathbf{b}^{(2)}$, the argument applies equally to any integrable two-particle interaction identified in  Section \ref{202pm81} or \ref{452pm81}.

The second difference lies in the form of the scattering matrix. Due to the change in the boundary condition, the expression  (\ref{623pm723}) is replaced by
\begin{equation}\label{623pm81}
\mathbf{R}_{\beta\alpha} = -(\mathbf{I}\otimes \mathbf{I} - \frac{1}{\xi_{\beta}}\mathbf{B})^{-1}(\mathbf{I}\otimes \mathbf{I} - \frac{1}{\xi_{\alpha}}\mathbf{B}),
\end{equation}
while the rest of the arguments, including the structure of the Bethe ansatz and Yang–Baxter equation, remain unchanged.
\subsection*{Transition probabilities}
Once integrability has been established, the derivation of the transition probability $P_{(Y,\nu)}(X,\pi;t)$ follows standard methods developed in earlier works such as \cite{Lee-2020,Lee-2024,Schutz-1997,Tracy-Widom-2008}.

Consider an $n$-particle system where each particle's species can be one of the integers $\{1,\dots, N\}$.  Let $\mathbf{B}$ be an integrable  $N^2 \times N^2$ matrix describing two-particle interactions, as discussed in Section \ref{202pm81} and \ref{452pm81}. As shown previously, the function
\begin{equation}\label{600pm610A}
\mathbf{U}(x_1,\dots,x_n;t) = \sum_{\sigma \in \mathcal{S}_n} \mathbf{A}_{\sigma}\prod_{i=1}^n\xi_{\sigma(i)}^{x_i} e^{\varepsilon(\xi_i)t},
\end{equation}
where $\mathbf{A}_{\sigma}$ is defined by (\ref{623pm723}),(\ref{803pm68}), and (\ref{232pm618}), satisfies the master equation that $\mathbf{P}_Y(X;t)$ satisfies for each  $X=(x_1,\dots,x_n)$ with $x_1< \cdots <x_n$. The initial condition is given by
\begin{equation*}
\mathbf{P}_Y(X;0) = \begin{cases}
\textrm{the identity matrix}&~\textrm{if $X = Y$} \\[4pt]
\textrm{the zero matrix}&~\textrm{if $X \neq Y$}.
\end{cases}
\end{equation*}
Then $\mathbf{P}_Y(X;t)$ admits the contour integral representation:
\begin{equation*}
\mathbf{P}_Y(X;t)=\dashint_c\cdots \dashint_c\sum_{\sigma\in \mathcal{S}_N}\mathbf{A}_{\sigma}\prod_{i=1}^N\Big(\xi_{\sigma(i)}^{x_i-y_{\sigma(i)}-1}e^{\varepsilon(\xi_i) t}\Big)~ d\xi_1\cdots d\xi_N,
\end{equation*}
where $c$ is a positively oriented circle centered at the origin with radius less than 1 and $\dashint=(1/2\pi i)\int$, and the integrals are evaluated matrix entry-wise.
\subsection*{Summary and outlook}
In this work, we have extended the classification of integrable two-species stochastic particle systems with homogeneous rates, originally developed by Alimohammadi and Ahmadi, to the general $N$-species setting. Among the 28 integrable two-species interaction matrices found by Alimohammadi and Ahmadi, we identified precisely 7 that admit consistent and natural extensions to integrable $N$-species models. Furthermore, we demonstrated that several convex linear combinations of these 7 matrices also yield integrable models, and we introduced a broader two-parameter family of interaction matrices that generalizes many of these cases.

To broaden the scope of integrability beyond the natural extension, we proposed an asymmetric extension scheme. This approach allows a distinguished species (e.g., species 1) to follow different interaction rules from the others. Using this method, we identified 8 additional two-species interaction matrices that can be extended to integrable $N$-species models. These findings reveal that the Yang-Baxter integrability can persist even under structured asymmetries in interaction rules.

We also showed that all integrable models identified in this paper retain their integrability under a change in dynamics from the backward-jumping rule (as in TASEP) to the forward-jumping rule (as in drop-push models). For all such models, we derived explicit transition probability formulas.

Looking ahead, several directions remain open. First, a complete classification of integrable $N$-species models beyond the natural extension  or the asymmetric extensions introduced in this work remains elusive. Second, it would be valuable to investigate whether these integrable models exhibit rich combinatorial structures similar to those found in the ASEP and PushTASEP, particularly in the context of computing one-point distributions as in  \cite{Tracy-Widom-2008} or  the multi-point distribution as in \cite{Borodin2}. Third, the study of asymptotic behavior and fluctuation properties, such as those predicted by the KPZ universality class, remains largely unexplored for the newly discovered integrable families, especially under special initial conditions. We hope that this work provides a foundation for further developments in the theory of integrable multispecies models and fosters deeper connections between integrable probability, combinatorics, and nonequilibrium statistical mechanics.

\textbf{Acknowledgement} This research is funded by Nazarbayev University under the faculty-development competitive research grants program
for 2024 -- 2026 (grant number 201223FD8822).
\\ \\
\textbf{Conflict of Interest} The author declares that there is no conflict of interest.
\newpage
\begin{appendix}
\section{Matrices for two-particle interactions in integrable ``two-species" models}\label{627pm729}
Alimohammadi and Ahmadi \cite{Ali-Ahmadi} classified all $4\times 4$ matrices $\mathbf{b}$  (up to the interchange symmetry between species 1 and 2) that satisfy  the following conditions.
\begin{itemize}
  \item [(i)] Each column of $\mathbf{b}$ sums to 1.
  \item [(ii)] All off-diagonal entries are either $0$ or $1$.
  \item [(iii)] The associated scattering matrices
\begin{equation*}
\mathbf{R}_{\beta\alpha} = -(\mathbf{I}\otimes \mathbf{I} - \xi_{\alpha}\mathbf{b})^{-1}(\mathbf{I}\otimes \mathbf{I} - \xi_{\beta}\mathbf{b})
\end{equation*}
where $\mathbf{I}$ denotes the $2 \times 2$ identity matrix, and $\xi_{\alpha},\xi_{\beta}$ are spectral parameters, satisfy the relation (Yang-Baxter equation) between $8 \times 8$ matrices:
\begin{equation*}
(\mathbf{R}_{\gamma\beta} \otimes \mathbf{I})(\mathbf{I} \otimes \mathbf{R}_{\gamma\alpha})(\mathbf{R}_{\beta\alpha} \otimes \mathbf{I}) = (\mathbf{I} \otimes \mathbf{R}_{\beta\alpha})(\mathbf{R}_{\gamma\alpha} \otimes \mathbf{I})(\mathbf{I} \otimes \mathbf{R}_{\gamma\beta}).
\end{equation*}
\end{itemize}
Their classification yields exactly the following 28 matrices. For brevity, the row and columns labels $11,12,21,22$ are  omitted in the following matrices.
\begin{equation*}
 \mathbf{b}^{(1)}=\left(
      \begin{array}{cccc}
        1 & 0 & 0 & 0 \\
        0 & 1 & 0 & 0 \\
        0 & 0 & 1 & 0 \\
        0 & 0 & 0 & 1 \\
      \end{array}
    \right), ~~~ \mathbf{b}^{(2)}=\left(
      \begin{array}{cccc}
        1 & 0 & 0 & 0 \\
        0 & 0 & 0 & 0 \\
        0 & 1 & 1 & 0 \\
        0 & 0 & 0 & 1 \\
      \end{array}
    \right),~~~ \mathbf{b}^{(3)}=\left(
      \begin{array}{cccc}
        1 & 0 & 0 & 0 \\
        0 & 1 & 0 & 0 \\
        0 & 0 & 0 & 0 \\
        0 & 0 & 1 & 1 \\
      \end{array}
    \right)
\end{equation*}
\begin{equation*}
 \mathbf{b}^{(4)}=\left(
      \begin{array}{cccc}
        1 & 0 & 0 & 0 \\
        0 & 0 & 0 & 0 \\
        0 & 0 & 1 & 0 \\
        0 & 1 & 0 & 1 \\
      \end{array}
    \right), ~~~ \mathbf{b}^{(5)}=\left(
      \begin{array}{cccc}
        1 & 0 & 0 & 0 \\
        0 & 0 & 0 & 0 \\
        0 & 0 & 0 & 0 \\
        0 & 1 & 1 & 1 \\
      \end{array}
    \right),~~~ \mathbf{b}^{(6)}=\left(
      \begin{array}{cccc}
        0 & 0 & 0 & 0 \\
        0 & 0 & 0 & 0 \\
        0 & 1 & 1 & 0 \\
        1 & 0 & 0 & 1 \\
      \end{array}
    \right)
\end{equation*}
\begin{equation*}
 \mathbf{b}^{(7)}=\left(
      \begin{array}{cccc}
        0 & 0 & 0 & 0 \\
        0 & 0 & 0 & 0 \\
        1 & 0 & 1 & 0 \\
        0 & 1 & 0 & 1 \\
      \end{array}
    \right), ~~~ \mathbf{b}^{(8)}=\left(
      \begin{array}{cccc}
        0 & 0 & 0 & 0 \\
        0 & 0 & 0 & 0 \\
        1 & 1 & 1 & 0 \\
        0 & 0 & 0 & 1 \\
      \end{array}
    \right),~~~ \mathbf{b}^{(9)}=\left(
      \begin{array}{cccc}
        0 & 0 & 0 & 0 \\
        0 & 1 & 1 & 0 \\
        0 &0  & 0 & 0 \\
        1 & 0 & 0 & 1 \\
      \end{array}
    \right)
\end{equation*}
\begin{equation*}
 \mathbf{b}^{(10)}=\left(
      \begin{array}{cccc}
        0 & 0 & 0 & 0 \\
        1 & 1 & 0 & 0 \\
        0 & 0 & 1 & 1 \\
        0 & 0 & 0 & 0 \\
      \end{array}
    \right), ~~~ \mathbf{b}^{(11)}=\left(
      \begin{array}{cccc}
        1 & 0 & 1 & 0 \\
        0 & 0 & 0 & 0 \\
        0 & 0 & 0 & 0 \\
        0 & 1 & 0 & 1 \\
      \end{array}
    \right),~~~ \mathbf{b}^{(12)}=\left(
      \begin{array}{cccc}
        0 & 0 & 0 & 0 \\
        1 & 1 & 1 & 0 \\
        0 & 0 & 0 & 0 \\
        0 & 0 & 0 & 1 \\
      \end{array}
    \right)
\end{equation*}
\begin{equation*}
 \mathbf{b}^{(13)}=\left(
      \begin{array}{cccc}
        1 & 1 & 0 & 0 \\
        0 & 0 & 0 & 0 \\
        0 & 0 & 0 & 0 \\
        0 & 0 & 1 & 1 \\
      \end{array}
    \right), ~~~ \mathbf{b}^{(14)}=\left(
      \begin{array}{cccc}
        0 & 0 & 0 & 0 \\
        1 & 1 & 0 & 0 \\
        0 & 0 & 0 & 0 \\
        0 & 0 & 1 & 1 \\
      \end{array}
    \right),~~~ \mathbf{b}^{(15)}=\left(
      \begin{array}{cccc}
        0 & 0 & 0 & 0 \\
        0 & 1 & 0 & 1 \\
        1 & 0 & 1 & 0 \\
        0 & 0 & 0 & 0 \\
      \end{array}
    \right)
\end{equation*}
\begin{equation*}
  \mathbf{b}^{(16)}=\left(
      \begin{array}{cccc}
        1 & 0 & 0 & 0 \\
        0 & -1 & 0 & 0 \\
        0 & 1 & 1 & 1 \\
        0 & 1 & 0 & 0 \\
      \end{array}
    \right), ~~~ \mathbf{b}^{(17)}=\left(
      \begin{array}{cccc}
        0 & 0 & 0 & 0 \\
        0 & 0 & 0 & 0 \\
        0 & 0 & 0 & 0 \\
        1 & 1 & 1 & 1 \\
      \end{array}
    \right),~~~ \mathbf{b}^{(18)}=\left(
      \begin{array}{cccc}
        0 & 0 & 0 & 0 \\
        1 & 0 & 1 & 0 \\
        0 & 0 & 0 & 0 \\
        0 & 1 & 0 & 1 \\
      \end{array}
    \right)
\end{equation*}
\begin{equation*}
  \mathbf{b}^{(19)}=\left(
      \begin{array}{cccc}
        -1 & 0 & 0 & 0 \\
        1 & 0 & 0 & 0 \\
        1 & 0 & 0 & 0 \\
        0 & 1 & 1 & 1 \\
      \end{array}
    \right), ~~~ \mathbf{b}^{(20)}=\left(
      \begin{array}{cccc}
        0 & 0 & 1 & 0 \\
        0 & 0 & 0 & 1 \\
        1 & 0 & 0 & 0 \\
        0 & 1 & 0 & 0 \\
      \end{array}
    \right),~~~ \mathbf{b}^{(21)}=\left(
      \begin{array}{cccc}
        0 & 1 & 0 & 0 \\
        1 & 0 & 0 & 0 \\
        0 & 0 & 0 & 1 \\
        0 & 0 & 1 & 0 \\
      \end{array}
    \right)
\end{equation*}
\begin{equation*}
  \mathbf{b}^{(22)}=\left(
      \begin{array}{cccc}
        0 & 0 & 0 & 1 \\
        0 & 0 & 1 & 0 \\
        0 & 1 & 0 & 0 \\
        1 & 0 & 0 & 0 \\
      \end{array}
    \right), ~~~ \mathbf{b}^{(23)}=\left(
      \begin{array}{cccc}
        0 & 0 & 1 & 0 \\
        1 & -1 & 1 & 0 \\
        0 & 1 & -1 & 1 \\
        0 & 1 & 0 & 0 \\
      \end{array}
    \right),~~~ \mathbf{b}^{(24)}=\left(
      \begin{array}{cccc}
        -1 & 1 & 0 & 1 \\
        0 & 0 & 0 & 1 \\
        1 & 0 & 0 & 0 \\
        1 & 0 & 1 & -1 \\
      \end{array}
    \right)
\end{equation*}
\begin{equation*}
 \mathbf{b}^{(25)}=\left(
      \begin{array}{cccc}
        -1 & 0 & 1 & 1 \\
        0 & -1 & 1 & 1 \\
        1 & 1 & -1 & 0 \\
        1 & 1 & 0 & -1\\
      \end{array}
    \right), ~~~ \mathbf{b}^{(26)}=\left(
      \begin{array}{cccc}
        -1 & 1 & 0 & 1 \\
        1 & -1 & 1 & 0 \\
        0 & 1 & -1 & 1 \\
        1 & 0 & 1 & -1 \\
      \end{array}
    \right)
\end{equation*}
\begin{equation*}
\mathbf{b}^{(27)}=\left(
      \begin{array}{cccc}
        -1 & 1 & 1 & 0 \\
        1 & -1 & 0 & 1 \\
        1 & 0 & -1 & 1 \\
        0 & 1 & 1 & -1 \\
      \end{array}
    \right), ~~~\mathbf{b}^{(28)}=\left(
      \begin{array}{cccc}
        -2 & 1 & 1 & 1 \\
        1 & -2 & 1 & 1 \\
        1 & 1 & -2 & 1 \\
        1 & 1 & 1 & -2 \\
      \end{array}
    \right)
\end{equation*}

\end{appendix}

\end{document}